\documentclass[aps, prl, preprint,english,superscriptaddress]{revtex4-1}

\usepackage{graphicx}
\usepackage{amsmath}
\usepackage{amssymb}
\usepackage{csquotes}
\MakeOuterQuote{"}
\usepackage{bm}
\usepackage{tikz}
\usepackage{tabularx}
\usepackage{hyperref}
\usepackage{color}
\usepackage[export]{adjustbox}

\begin{document}

\title{Role of kinetic instability in runaway electron avalanche and elevated critical electric fields}

\author{Chang Liu}
\affiliation{Princeton Plasma Physics Laboratory, Princeton, New Jersey 08540, USA}
\author{Eero Hirvijoki}
\affiliation{Princeton Plasma Physics Laboratory, Princeton, New Jersey 08540, USA}
\author{Guo-yong Fu}
\affiliation{Zhejiang University, Hangzhou, Zhejiang, 310027, China}
\affiliation{Princeton Plasma Physics Laboratory, Princeton, New Jersey 08540, USA}
\author{Dylan P. Brennan}
\affiliation{Princeton University, Princeton, New Jersey 08544, USA}
\author{Amitava Bhattacharjee}
\affiliation{Princeton Plasma Physics Laboratory, Princeton, New Jersey 08540, USA}
\affiliation{Princeton University, Princeton, New Jersey 08544, USA}
\author{Carlos Paz-Soldan}
\affiliation{General Atomics, San Diego, California 92186, USA}

\date{\today}

\begin{abstract}

The effects of kinetic whistler-wave instabilities on the runaway-electron (RE) avalanche is investigated. With parameters from DIII-D experiments, we show that RE scattering from excited whistler waves can explain several poorly understood experimental results seen in a variety of tokamaks. We find an increase of the avalanche growth rate and threshold electric field, bringing the present model much closer to observations than previous results. The excitation of kinetic instabilities and the scattering of resonant electrons are calculated self-consistently using a quasilinear model. We also explain the observed fast growth of electron cyclotron emission (ECE) signals and excitation of very low-frequency whistler modes observed in the quiescent RE experiments at DIII-D [D. A. Spong et al., submitted to Phys. Rev. Lett.]. These results indicate that by controlling the background thermal plasma temperature, the plasma wave can be excited spontaneously in tokamak disruptions and the avalanche generation of runaway electrons may be suppressed. 

\end{abstract}

\maketitle

\textit{Introduction.}-
In a plasma, the collisional friction force decreases with increasing electron velocity and a strong electric field acceleration force can overcome the collisional damping, accelerating high energy electrons to relativistic speeds. Such electrons are referred to as "runaway electrons" (REs). In tokamaks, REs have attracted a lot of attention due to their deleterious effects on the experimental device during disruption events\cite{knoepfel_runaway_1979,boozer_theory_2015}. Studies of RE dynamics reveal that knock-on collisions can lead to avalanche multiplication of the RE population\cite{rosenbluth_theory_1997}, and calculations show that the threshold electric field for this scenario, the Connor-Hastie electric field $E_{\mathrm{CH}}$\cite{connor_relativistic_1975}, is much smaller than the typical induced electric field in disruptions. Consequently, a strong RE avalanche effect could convert a large fraction of the initially Ohmic plasma current to RE current during disruptions. Aside from tokamak plasmas, RE avalanche is also important also in other areas, such as lightning formation during thunderstorms\cite{gurevich_runaway_1992}.

Quiescent runaway electron experiments in well-controlled stable scenarios (called flattop) with low electron density have been conducted in several tokamaks\cite{fussmann_long-pulse_1981,paz-soldan_growth_2014,zhou_investigation_2013}. The RE density and energy distributions have been inferred from radiation signals, including hard X-rays (HXR), gamma rays and electron cyclotron emission (ECE). An important and surprising finding in these experiments has been that the value of the threshold electric field for the RE population to transition from growth to decay is not the expected $E_{\mathrm{CH}}$\cite{granetz_itpa_2014,paz-soldan_growth_2014} but 5-10 times higher. This discrepancy with the avalanche theory indicates the presence of anomalous RE loss mechanisms, and numerical simulations have indeed indicated the importance of radiative energy losses for highly energetic REs, including synchrotron\cite{andersson_damping_2001,stahl_effective_2015} and bremsstrahlung losses\cite{bakhtiari_role_2005,embreus_effect_2016}. However, theoretical calculations taking these effects into account raise the threshold electric field to about 2 $E_{CH}$, which is still much smaller than that observed. In recent DIII-D experiments\cite{paz-soldan_spatiotemporal_2017}, the gamma-ray imaging (GRI) shows that the RE density decreases in the low energy regime while $E$ remains several times $E_{\mathrm{CH}}$, which differs from numerical simulation results including radiative losses. In addition, during both flattop\cite{paz-soldan_non-thermal_2016} and the RE plateau in disruption experiments\cite{fredrickson_control_2015}, strong ECE from REs was observed. These observations are suggestive of strong pitch angle scattering in the RE population at low energy, which could significantly enhance radiative losses.


In this Letter, we demonstrate that a self-consistent kinetic treatment which includes the whistler-wave instability driven by the highly anisotropic RE momentum-space distribution 
provides a conclusive resolution of the observed discrepancy between theory and experiment. New simulations presented in this paper produce a threshold of $6.4 E_\mathrm{CH}$, which is very close to observations. 

Previous studies of kinetic instabilities associated with a RE\cite{parail_kinetic_1978} have addressed the instability criterion\cite{fulop_destabilization_2006}, the diffusion effect from on the electron distribution\cite{pokol_quasi-linear_2008,komar_interaction_2012,pokol_quasi-linear_2014,komar_electromagnetic_2013}, and the convective aspect of the instabilities\cite{aleynikov_stability_2015}. A detailed study containing also the avalanche effect and radiation damping is, however, missing. In the present work, we present a recently developed numerical model to study the evolution of both the runaway electron distribution and the wave energy spectrum self-consistently, including the avalanche source term from knock-on collisions and the radiation reaction. The model includes the wave-particle interaction within the quasilinear approximation. 
Our numerical analysis of a typical low-density DIII-D experiment reveals that in addition to  the low frequency whistler waves (LFWWs), the high frequency whistler waves (HFWWs, the whistler waves at high frequency end near the resonant cone) can also be excited and scatter runaway electrons in the low energy regime.
The results show an increase of the critical electric field and a rapid increase in the ECE emission from REs, also seen in the experiments.

\textit{Simulation framework.}-
The whistler wave belongs to the fast wave branch of the plasma wave dispersion relation.
In this work, the frequency and the polarization of the whistler waves for every $(k,\theta)$ (with $\theta=\arccos k_{\parallel}/k$) are calculated using the cold plasma dielectric tensor. We also calculate the collisional damping rate of every mode according to the electron-ion collisional frequency\cite{aleynikov_stability_2015}.

The evolution of the electron distribution function $f$ in momentum space is advanced through the kinetic equation. The coordinates for momentum space are $(p,\xi)$, where $p$ is the electron momentum normalized to $mc$ ($m$ is the electron mass and $c$ is the speed of light), and $\xi=p_{\parallel}/p$ is the cosine of the pitch angle. The kinetic equation we solve is
\begin{equation}
\frac{\partial f}{\partial t}+\frac{eE_{\parallel}}{mc}\left(\xi \frac{\partial f}{\partial p}+\frac{1-\xi^2}{p}\frac{\partial f}{\partial \xi}\right)+C\left[f\right]+\frac{\partial}{\partial \mathbf{p}}\cdot\left(\mathbf{F}_{\mathrm{rad}}f\right)+D\left[f\right]=S_{A}\left[f\right],
\end{equation}
with $E_{\parallel}$ the parallel electric field, 
$C[\dots]$ the test-particle collision operator\cite{landreman_numerical_2014}. $\mathbf{F}_{\mathrm{rad}}$ is the synchrotron radiation reaction force\cite{stahl_effective_2015}. $D\left[\dots\right]$ is the diffusion operator from the excited waves. $S_{A}\left[\dots\right]$ is the source term for the runaway electron avalanche\cite{liu_adjoint_2017}. 

Given the distribution function, we can obtain the growth (or damping) rate $\Gamma$ of every mode, using\cite{aleynikov_stability_2015}
\begin{equation}
\label{eq:growth-rate}
\Gamma(k,\theta)  = \frac{{\omega _{pe}^2}}{\mathcal{D}}\int {{d^3}p\mathop \sum \limits_{n =  - \infty }^{n = \infty } {Q_n}\pi \delta (\omega  - {k_\parallel }v\xi  - n{\omega _{ce}}/\gamma )(p^{2}/\gamma)\hat L f},
\end{equation}
where
\begin{equation}
{Q_n} = {\left[ {\frac{{n{\omega _{ce}}}}{{\gamma {k_ \bot }v}}{J_n}({k_ \bot }\rho ) + {E_z}\xi {J_n}({k_ \bot }\rho ) + i{E_y}\sqrt {1 - {\xi ^2}} J{'_n}({k_ \bot }\rho )} \right]^2},
\end{equation}
\begin{equation}
\hat L =  {\frac{1}{p}\frac{{\partial}}{{\partial p}} - \frac{1}{p^{2}}\frac{{n{\omega _{ce}}/\gamma  - \omega (1 - {\xi ^2})}}{{\omega \xi }}\frac{{\partial }}{{\partial \xi }}},
\end{equation}
Here $\omega_{pe}$ and $\omega_{ce}$ are the plasma frequency and electron cyclotron frequency (we choose $\omega_{ce}<0$), $J_{n}$ is the $n$th order Bessel function, $v$ is the particle velocity, 
$\gamma$ is the relativistic factor, and $\rho=m p\sqrt{1-\xi^2}/\omega_{ce}$ is the Larmor radius. $\mathcal{D}$ is from Eq. (21) in \cite{aleynikov_stability_2015}. $E_{y}$ and $E_{z}$ are wave polarization normalized to $E_{x}$. $f$ is normalized so that $\int p^{2}dp d\xi f=1$.

The wave energy $\mathcal{E}(k,\theta)$ then evolves as
\begin{equation}
\label{eq:E-evolution}
\frac{d \mathcal{E}(k,\theta)^{2}}{d t}=2\Gamma(k,\theta) \mathcal{E}(k,\theta)^{2}+\mathcal{K}(k,\theta),
\end{equation}
where $\mathcal{K}(k,\theta)$ represents the background fluctuation electromagnetic field energy from radiations, which provides the initial amplitudes of the modes\cite{harvey_electron_1993}. $\mathcal{K}$ can be calculated as
\begin{equation}
\label{eq:E-emission}
\mathcal{K}(k,\theta)  = \frac{{\omega _{pe}^2}}{\mathcal{D}}\int {{d^3}p\mathop \sum \limits_{n =  - \infty }^{n = \infty } {Q_n}\pi \delta (\omega  - {k_\parallel }v\xi  - n{\omega _{ce}}/\gamma ) mv^{2} f}.
\end{equation}
The diffusion of resonant electrons in momentum space can be calculated using the wave energy in a quasilinear diffusion model\cite{kaufman_quasilinear_1972},
\begin{equation}
\label{eq:quasilinear-diffusion}
D[f] = \frac{e^2}{2\mathcal{D}}\sum\limits_{n =  - \infty }^\infty  {\int {{d^3}{\bf{k}}\,\hat L \left[{p_ \bot }\delta (\omega  - {k_\parallel }{v \xi } - n\omega_{ce}/\gamma )\mathcal{E}(k,\theta){Q_n}{p_ \bot }\hat L {f }\right]} }.
\end{equation}

Note that in Eqs. (\ref{eq:growth-rate}), (\ref{eq:E-emission}) and (\ref{eq:quasilinear-diffusion}), the wave-particle interaction happens only when the resonance condition is satisfied: $\omega-k_{\parallel} v\xi=n\omega_{ce}/\gamma$. This includes Cherenkov resonance ($n=0$), normal Doppler resonance ($n<0$) and anomalous Doppler resonance ($n>0$).
For $n<0$, the resonant momentum $p$ is a decaying function of  $\omega$ for a fixed $\theta$, thus the low energy electron will resonate with high frequency waves and vice versa. For Cherenkov resonance, $p$ is a non-monotonic function of $k$ and $\theta$, so for LFWWs and HFWWs, the resonance regions overlap.


The numerical representation of $f$ and $\mathcal{E}$ is adjusted, guided by the anticipated shape of the solution. For $f$, we use the finite element method with 1000 elements in $p$ and 50 elements in $\xi$.
For the wave energy spectrum, we use a mesh with 50 points in $\theta$ and $160$ points in $k$. For every mode, we calculate a line integral for $f$ to obtain $\Gamma$ and $\mathcal{K}$ according to Eq. (\ref{eq:growth-rate})(\ref{eq:E-emission}). For the diffusion operator, we use linear interpolation to get the wave energy required by Eq. (\ref{eq:quasilinear-diffusion}) for every quadrature point in the $f$ mesh. In the calculation of Eq. (\ref{eq:growth-rate}) and Eq. (\ref{eq:quasilinear-diffusion}), we only include $n=0, \pm 1$ assuming they are the most dominant resonances. The timestep is chosen according to $\Delta t=1/\Gamma_{\mathrm{max}}$, where $\Gamma_{\mathrm{max}}$ is the maximum  value of $\Gamma$ for all the modes. For every timestep, the evolution of $\mathcal{E}$ is calculated by integrating \ref{eq:E-evolution} for $\Delta t$. 

To better validate against experiments, we also developed an ECE synthetic diagnostic code to calculate the ECE radiation power from the electron distribution\cite{harvey_electron_1993,liu_runaway_2017}. Note that in the current model we only have the electron distribution in 2D momentum space, thus to calculate the ECE signals, we assume that the electron distribution containing runaway tail is uniform near the core from $-0.5a$ to $0.5a$ ($a$ is the minor radius), and outside this region, the electron distribution is assumed to be a Maxwellian distribution with a specified temperature profile.


\textit{Simulation of flattop RE experiment scenarios.}-
We now use the model to simulate a DIII-D flattop experiment, which has strong runaway electron generation due to the avalanche process. Note that a typical RE discharge consists of two stages. In the first stage, the plasma density is very low and the parallel electric field supporting the Ohmic current is sufficient to accelerate a RE tail. When the REs reach a critical intensity, an asynchronous trigger begins the RE dissipation stage, in which the electron density is varied by gas puffing. The parameters we use in simulation are close to the numbers from the tokamak core diagnostic. For stage 1, $n_{e}=0.6\times 10^{19}$m$^{-3}$, $T_{e}=1.3$keV,
and $B=1.45$T. $E_{\parallel}=0.055$V/m, which is about $9E_{\mathrm{CH}}$. For stage 2, we increase the density to $n_{e}=0.8\times 10^{19}$, by adding a Maxwellian part to $f$ from the last timestep of stage 1. We also decrease the electric field to $E=0.045$V/m, so $E/E_{\mathrm{CH}}$ becomes 5.5.

\begin{figure}[h]
	\begin{center}
		\includegraphics[width=0.9\linewidth]{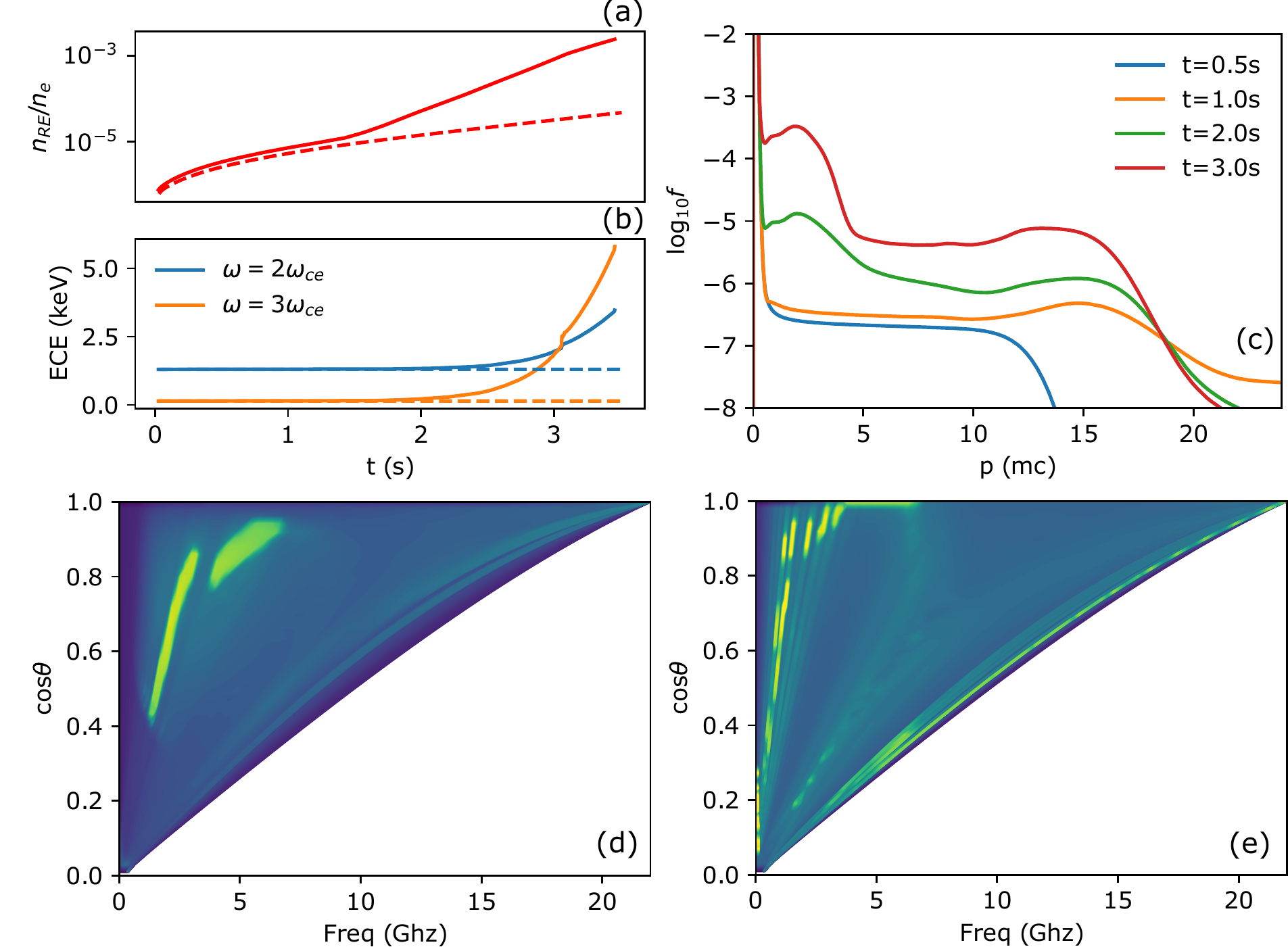}
	\end{center}
	\caption{\label{fig:before}(a) Growth of RE density
	 with time, with wave diffusion (solid) and without (dashed). (b) ECE signals of second and third core $\omega_{ce}$ from synthetic diagnostics, with wave diffusion (solid) and without (dashed). (c) Evolution of $f$ integrated over the pitch angle. (d) Whistler wave energy spectrum for $t=1.0$s. Right boundary is the whistler wave resonance cone. (e) Whistler wave energy spectrum for $t=3.0$s.}
\end{figure}

The simulation result of stage 1 is summarized in Fig. \ref{fig:before}. In the early time (before 0.5s), the runaway electron tail is formed through Dreicer generation (Fig. \ref{fig:before} (c)). In this case all the modes are stable. Then as the runaway electron tail extends to $p=15$, the LFWWs (from 1GHz to 5GHz, as shown in Fig. \ref{fig:before}(d)) first get excited. This gives rise to strong pitch angle scattering for high energy runaway electrons
(as shown below in Fig. \ref{fig:f2d}(a)). This effect is known as the "fan instability" and has been studied previously\cite{parail_kinetic_1978,pokol_quasi-linear_2008}.

\begin{figure}[h]
	\begin{center}
		\includegraphics[width=0.8\linewidth]{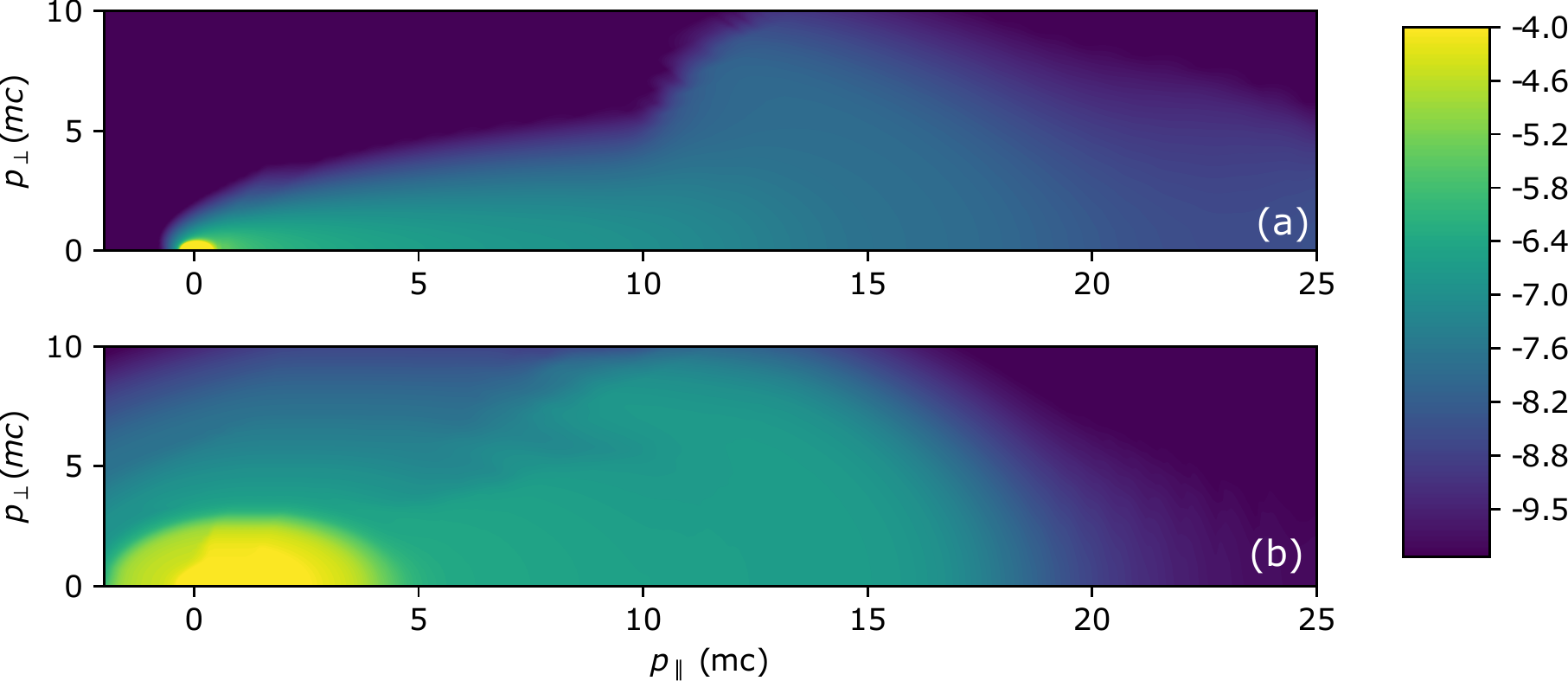}
	\end{center}
	\caption{\label{fig:f2d} (a) Value of $\log_{10}f$ in $p_{\parallel}-p_{\perp}$ space at $t=1.0$s. (b) Value of $\log_{10}f$ in at $t=3.0$s.}
\end{figure}

After 1.6s, the HFWWs also get excited as shown in Fig. \ref{fig:before} (e) (the wave close to the spectrum right boundary which is the resonance cone). These waves can resonate with lower energy electrons through anomalous Doppler resonance. The result of the excitation, as shown in Fig. \ref{fig:f2d}(b), is that the low energy runaway electrons can be scattered to very large pitch angle. This effect also leads to the fast growth of the ECE signals as shown in Fig. \ref{fig:before} (a). Calculation of ECE weight function\cite{liu_runaway_2017} shows that electrons in the low energy regime with large pitch angle are the most efficient at generating ECE power. This explains why the ECE signals only start to grow after the HFWWs get excited. We also observe that as the ECE signals grow, the higher frequency signal corresponding to the third harmonic surpasses the second harmonic, which agrees with the experiment observations\cite{paz-soldan_non-thermal_2016}.

As shown in Fig. \ref{fig:before} (b), after the whistler waves are excited, the avalanche growth rate increases. This effect is caused by diffusion of electrons in momentum space through Cherenkov resonance. Using the resonance condition, we find that the Cherenkov resonance region for the excited waves is about $0.3<v_{\parallel}/c<0.5$, which is close to the runaway-loss separatrix\cite{liu_adjoint_2016}. Note that the electron distribution function close to the separatrix satisfies $\partial f/\partial p_{\parallel}<0$, so diffusion makes low energy electrons move to higher energy and gain energy from the wave (Landau damping), and become more probable to runaway.

\begin{figure}[h]
    \begin{center}
			\includegraphics[width=0.9\linewidth]{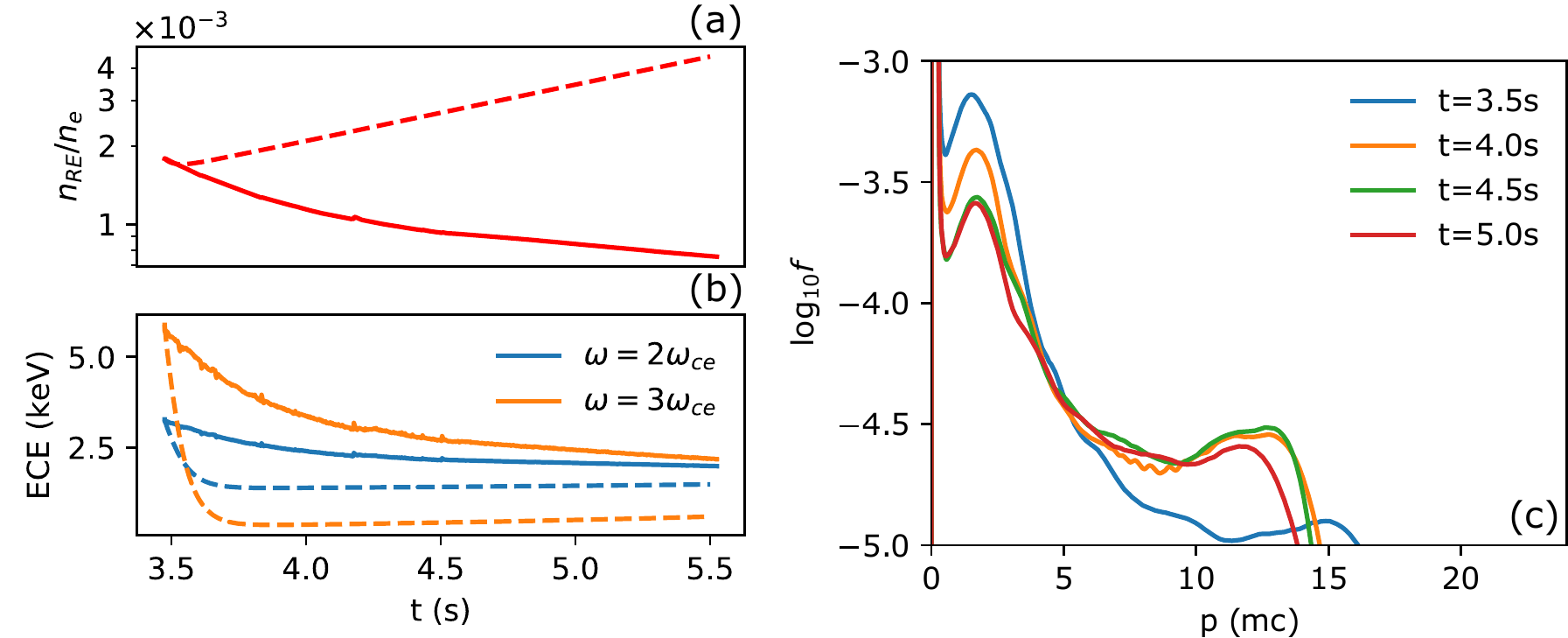}
    \end{center}
    \caption{\label{fig:after}(a) ECE signals from synthetic diagnostic, with wave diffusion (solid) and without (dashed). (b) Evolution of RE density, with wave (solid) and without. (c) Evolution of $f$ integrated over pitch angle. }
\end{figure}

We now look at the simulation result in stage 2, which is shown in Fig. \ref{fig:after}. According to the previous theories of the effective critical electric field\cite{stahl_effective_2015,aleynikov_theory_2015}, the electric field is still larger than the critical electric field including the synchrotron radiation loss, so the runaway population would grow if this were the extent of the model. This is confirmed in our simulation without wave diffusion, as shown by the dashed line in Fig. \ref{fig:after} (a). However, with the wave diffusion, we find that the runaway electron population actually starts to decay under these conditions. Examination of the evolution of $f$ (Fig. \ref{fig:after} (c)) reveals that this is mainly caused by the loss of the runaway electron population in the lower energy regime, which is in agreement with recent findings in DIII-D experiments\cite{paz-soldan_spatiotemporal_2017}.
The whistler wave spectrum and the shape of RE distribution function in stage 2 is similar to that of the later phase of stage 1.

The fact that the runaway population decays because of wave diffusion seemingly contradicts the results from stage 1. The reason is that, the wave diffusion not only can enhance the runaway avalanche through diffusion at the separatrix, but also provides a new mechanism of runaway electron loss. The diffusion effects from HFWWs can scatter REs to large pitch angle in the low energy regime. Given that electrons with large pitch angle are less susceptible to the electric force acceleration,
they more easily lose energy and return to the thermal population. In tokamaks this effect can be further enhanced by the inhomogeneity of magnetic field, since electrons with large pitch angle become trapped electrons and will not be further accelerated\cite{nilsson_trapped-electron_2015}.

\begin{figure}[h]
	\begin{center}
		\includegraphics[width=0.9\linewidth]{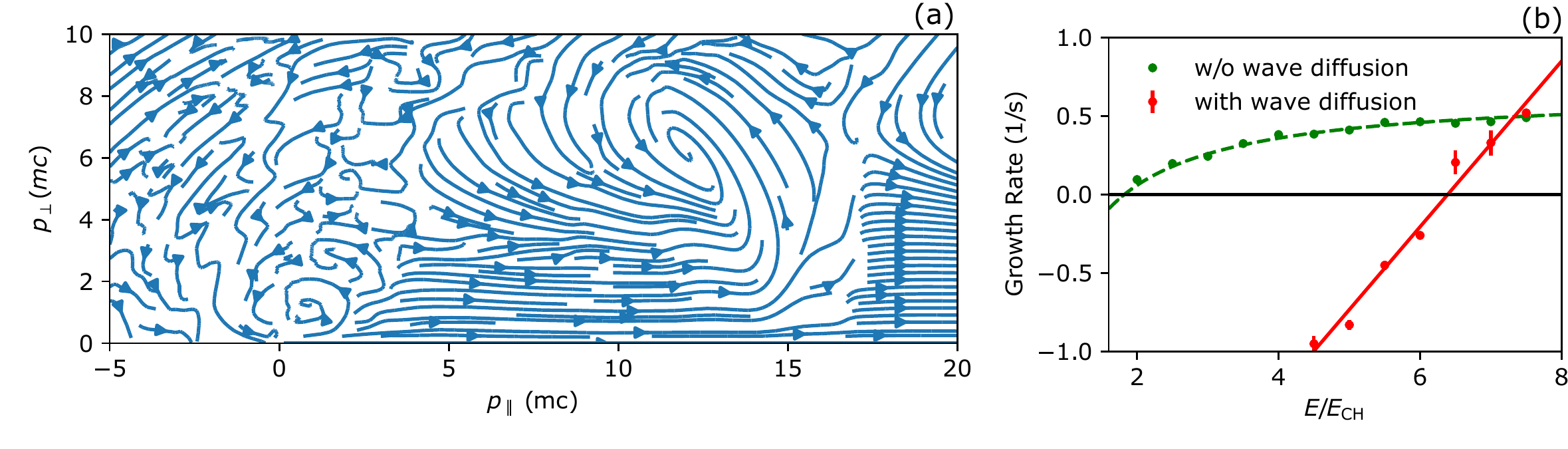}
	\end{center}
    \caption{\label{fig:vortex} (a) Electron flux in momentum space at $t=4.0$s including the whistler wave diffusion. (b) RE density growth (decay) rate as a function of $E/E_{\mathrm{CH}}$. The Red dots are from the simulation results with wave diffusion, and the red line is a linear regression. The green dots are from the simulation results without wave diffusion, and the green dashed line is the growth rate calculated from \cite{rosenbluth_theory_1997}, with critical electric field $E_\mathrm{c}=1.82 E_{\mathrm{CH}}$ from \cite{liu_adjoint_2017}}
\end{figure}

The effect of wave diffusion on RE dynamics in momentum space is further illustrated in Fig. \ref{fig:vortex} (a), where we show the directions of electron flux in momentum space calculated from the kinetic equation. In the high energy regime, a vortex structure is formed in the momentum space ($10<p_\parallel<15$) due to LFWWs. The location of the vortex is in a much lower energy regime than that from radiation forces\cite{guo_phase-space_2017}. This vortex can hinder REs from going into the higher energy regime, resulting in a bump-on-tail distribution. On the other hand, in the low energy regime ($-5<p_\parallel<3$) electron flux is stochastic since the dynamics are dominated by diffusion rather than advection. The strong diffusion in this region comes from both LFWWs and HFWWs through all 3 resonances. Electrons entering this region from the avalanche can be diffused from low pitch angle to high pitch angle, losing energy to the waves, and finally returning to the bulk electron population.

Combining the two effects, we find that the excitation of whistler waves can increase both the avalanche growth rate and the critical electric field. Scanning $n_e$ keeping other parameters fixed shows that the new critical electric field
is about $E/E_{\mathrm{CH}}=6.4$ as indicated in Fig. \ref{fig:vortex} (b). This value is much larger than the previous predictions without kinetic instabilities(Green line in Fig. \ref{fig:vortex} (b)), and is closer to the experimental observation. However, this value depends on plasma parameters including $T_{e}$, $n_{e}$, $B$, and the whistler wave amplitudes.

We also use our numerical model to study the excitation of very low frequency whistler waves, an effect that was recently observed\cite{spong_first_nodate} for the first time. By examining the whistler wave spectrum in stage 2, we find that a branch of whistler modes with frequency between 100MHz to 200MHz and $k_{\perp}\gg k_{\parallel}$ are excited in our simulation of stage 2.
Unlike previously discussed whistler waves, these waves are mainly driven by the Cherenkov resonance. The reason is that, the low energy runaway electrons scattered by the HFWW can accumulate in a certain region of momentum space, which can cause an unstable bump-on-tail distribution function and drive the very low frequency whistler waves. This is illustrated in Fig. \ref{fig:vortex} (a), where we see electrons above the vortex loosing $p_\parallel$ due to interaction with waves. 

Note that in the analysis of kinetic instabilities presented so far, we didn't take into account for the finite spatial extent of excited waves, termed the convective effect\cite{aleynikov_stability_2015}.
To understand it, we use a ray tracing code GENRAY\cite{smirnov_general_1994} to study the propagation of the whistler waves, launching from
the magnetic axis. We find that, for both LFWW and HFWW excited in our simulations,
the rays will have trajectories bouncing back and forth in the poloidal plane, and the values of $n_{\parallel}$ and $n_{\perp}$ will stay close to their initial values. This means that the convective stability of these waves will not be much different from the local analysis.

In addition to flattop experiments, we also study the excitation of kinetic instabilities in disruptions. However, we find that for typical parameters in post-disruption plateau in DIII-D, kinetic instabilities can hardly be excited due to strong collisional damping at very low temperature ($\sim 5$eV). In addition, the excited waves can only affect REs in the high energy regime, and have little effect on the avalanche. This is partly because the HFWW are more affected by the collisional damping than the LFWW\cite{aleynikov_stability_2015}. On the other hand, if we raise $T_e$ to about 50eV, we find that both LFWW and HFWW can be excited. This temperature is consistent with previous estimation of kinetic instability threshold\cite{aleynikov_stability_2015}. Note that in disruption experiments with possible higher temperature\cite{fredrickson_control_2015}, signals of kinetic instabilities have been observed. This means that in order to have the whistler waves excited and help mitigate the avalanche, it may be beneficial to increase the thermal plasma temperature using external heating techniques.

\textit{Summary.}-
To conclude, with the help of a newly-developed simulation model, we have advanced our understanding of the interactions of kinetic instabilities and the runaway electron avalanche and explained several outstanding experimental observations. We find that the excited kinetic instabilities can either enhance or suppress the runaway electron growth. The RE distribution function in momentum space, taking into account the wave diffusion, differs significantly from the classical runaway electron tail, inspiring a revisit of previous studies on RE. Using this model, we successfully explain several phenomena in DIII-D flattop RE experiments, including 1) the increase of the critical electric field, 2) the decaying of RE density in the low energy regime, 3) the ECE from runaway electrons, 4) the observation of very low frequency whistler waves.
These results suggest the possibility of controlling the runaway electron avalanche through kinetic instabilities, including both self-generation and launching waves externally.
\begin{acknowledgments}
Chang Liu would like to to thank T\"{u}nde F\"{u}l\"{o}p, Gergo Pokol, Allen Boozer, Boris Breizman, Lei Shi, Nicola Bertelli, Richard Harvey, Donald Spong and William Heidbrink for fruitful discussions. This work has received funding from
the Department of Energy 
under Grant No. DE-SC0016268 and DE-AC02-09CH11466.
\end{acknowledgments}
\bibliography{whistler-avalanche}
		
\end{document}